\documentclass{article}
\usepackage{cite}
\usepackage{amsmath,amssymb,amsfonts}
\usepackage{algorithmic}
\usepackage{graphicx,subfigure}
\usepackage{textcomp}
\usepackage{xcolor}
\begin{document}
\title{A Semantic Approach for User-Brand Targeting in On-Line Social Networks}
\author{Mariella Bonomo, Gaspare Ciaccio, Andrea De Salve, Simona E. Rombo}
\date{Department of Mathematics and Computer Science \\
University of Palermo}
\maketitle
\begin{abstract}
We propose a general framework for the recommendation of possible customers (users) to advertisers (e.g., brands) based on the comparison between On-line Social Network profiles. In particular, we represent both user and brand profiles as trees where nodes correspond to categories and sub-categories in the associated On-line Social Network. When categories involve posts and comments, the comparison is based on word embedding, and this allows to take into account the similarity between topics popular in the brand profile and user preferences. Results on real datasets show that our approach is successfull in identifying the most suitable set of users to be used as target for a given advertisement campaign.
\end{abstract}


\section{Introduction}

Social media have achieved a growing popularity, especially On-line Social Networks (OSNs) that enable users to present
themselves, discuss about their preferred topics, establish and maintain social connections with
others. This leads advertisers to
invest more effort into communicating with consumers
through on-line social networking, which  provides suitable platforms for
advertising and marketing. 

An important issue is how to optimize the effects of marketing communication in the context of OSNs. In particular, advertisers aim to involve in their campaigns those users who are the most likely interested ones. Automatic systems able to suggest a set of target users for advertising campaigns provide three main benefits: {\it (i)} minimization of costs for the dissemination of the advertising campaign through social media, which is often very expensive; {\it (ii)} improvement of the user experience in OSNs, since only the possibly interested customers are contacted with advertisements which could be useful for them; {\it (iii)} avoid the spread of unuseful information through OSNs.

We propose a general framework for the recommendation of the best possible consumers to be suggested as target for a specific advertisement campaign. The recommendation is based on the comparison between the OSNs profiles associated to users (possible customers) and advertisers (e.g., brands), according to the considered campaign. The OSNs profiles are modelled as trees, which hierarchically represent the relationships among categories and sub-categories on the OSN (e.g., gender, job, posts, etc.). Profile matching is then applied relying on such a tree representation, and suitable similarity measures are considered for each category. When categories involve textual documents containing information on interests and preferences (e.g., posts and comments), the document is represented as a bag of words and the Term Frequency-Inverse Document Frequency (TF-IDF) is used to weight the importance of the words inside the text.

We present some preliminary tests performed on real datasets, in order to show how our framework works on specific advertisement campaigns. The obtained results are promising, indeed the approach allows to correctly identify the most suitable users in all the considered situations.

\section{Related work}
Modeling the user profiles from social media raw data is usually a challenging task. The approaches proposed in the Literature to this aim may be roughly classified in two main categories. The first category includes approaches based on the analysis of user generated contents (here referred to as {\it semantic approaches}). As for the approaches in the second category, individuals are characterized by ''actions'', e.g., visited web pages ({\it action-based approaches}). Our framework belongs to the first category, although to the best of our knowledge the other techniques applied to brand-affinity matching are action-based approaches. Moreover, it is the first approach that compares user and brand profiles based on word embedding from user posts.

\paragraph{Semantic approaches}
The authors of \cite{Schwartz2013} use Differential Language Analysis (DLA) in order to find language features across millions of Facebook messages that distinguish demographic and psychological attributes. They show that their approach can yield additional insights (correlations between personality and behavior as manifest through language) and more information (as measured through predictive accuracy) than traditional a priori word-category approaches. 

The framework proposed in \cite{Lin2014} relies on a semi-supervised topic model to construct a representation of an app's version as a set of latent topics from version metadata and textual descriptions. The authors discriminate the topics based on genre information and weight them on a per-user basis, in order to generate a version-sensitive ranked list of apps for a target user.

In \cite{LiangZRK18} the authors propose a dynamic user and word embedding algorithm that
can jointly and dynamically model user and word representations in the same semantic space. They consider the context of streams of documents
in Twitter, and propose a scalable black-box variational inference algorithm
to infer the dynamic embeddings of both users and words in streams. They also propose a streaming keyword diversification model to diversify top-K keywords for characterizing users’ profiles over time.

\paragraph{Action-based approaches}
In \cite{Provost2009} individuals are associated each other due to some actions they share (e.g., they have visited the same web pages). The proximity between individuals on networks built upon such relationships is informative about their profile matching. In particular, brand-affinity audiences are built by selecting the social-network neighbors of existing brand actors, identified via co-visitation of social-networking pages. This is achieved without saving any information about the identities of the browsers or content of the social-network pages, thus allowing for user anonymization.

In \cite{Ahmed2011} compact and effective user profiles are generated from the history of user actions, i.e., a mixture of user interests over a period of time. The authors propose a streaming, distributed inference algorithm which is able to handle tens of millions of users. They show that their model contributes towards improved behavioral targeting of display advertising relative to baseline models that do not incorporate topical and/or temporal dependencies. 

 In \cite{Iglesias2011} a computer user behavior is represented as the sequence of the commands she/he types during her/his work. This sequence is transformed into a distribution of relevant subsequences of commands in order to find out a profile that defines its behavior. Also, because a user profile is not necessarily fixed but rather it evolves/changes, the authors propose an evolving method to keep up to date the created profiles using an Evolving Systems approach.

The observation that behavior of users is highly influenced by the behavior of their neighbors or community members is used in  \cite{XieLMLCR14} to enrich user profiles, based on latent user communities in collaborative tagging.

\section{Proposed Approach} 
\subsection{Basics}
\label{sub::basics}
\begin{figure}[tb]
\centering
\includegraphics[width=\textwidth]{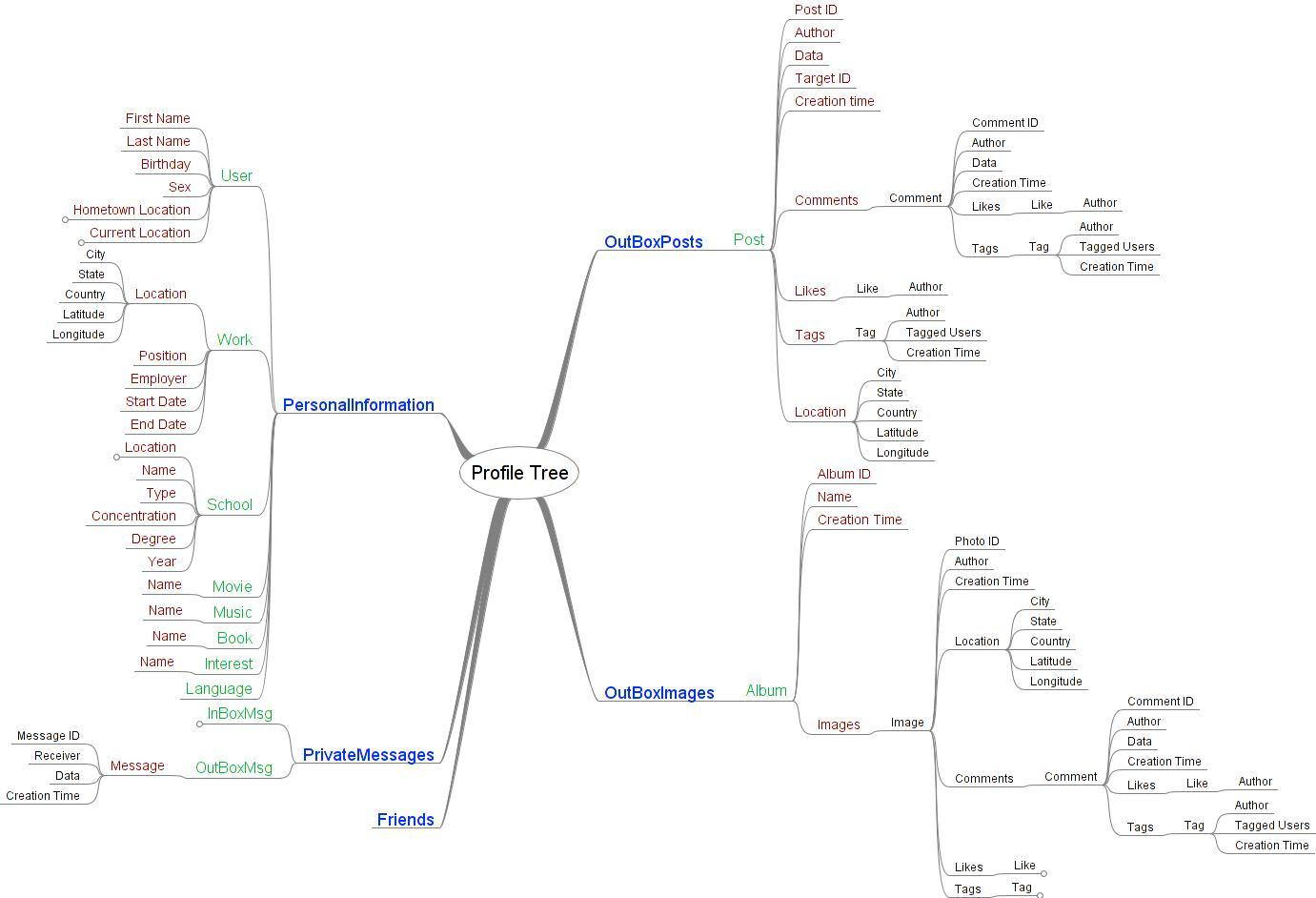}
\caption{General profile information of Facebook users}
\label{fig:profile}
\end{figure}

We represent the {\it social profile} (or profile) of a user $u$ in the OSN by a tree $P_u$ whose nodes correspond to categories and sub-categories of contents associated to that user in the OSN (see Figure \ref{fig:profile} for an illustrative example). The root of the tree
is considered as an entry point for all the categories of contents related to a profile owner, such as personal information, interests, friendship information, private communication, posts, images, comments, etc.. Each son $n_i$ of the root is associated to a standard category of contents in the OSN; the sub-tree rooted by $n_i$ hierarchically represents all the possible sub-categories of the $i$-th category and the leaves are associated to the instances of such sub-categories for $u$.

A basic notion for the research presented here is the {\it match} between nodes. We assume that all the non-leaves nodes match only if they are identical. Indeed, they represent categories and sub-categories, and we apply suitable data normalization at the first steps of our pipeline. Therefore, specific similarity functions and corresponding thresholds will be defined for the instances of the sub-categories, corresponding to leaves on the trees. In particular, let $P_u$ and $P_w$ be two profiles, $k$ be the number of possible categories, $\{\mathcal{S}_1, \ldots, \mathcal{S}_k\}$ be a set of similarity functions defined for each category $i\in\{1, \ldots, k\}$, and $\{th_1, \ldots, th_k\}$ be a set of thresholds for the values returned by those similarity functions. Given two leaves $l_u$ and $l_w$ which represent instances of the same sub-category in $P_u$ and $P_w$, respectively, they match if $\mathcal{S}_i(l_u,l_w) \geq th_i$, and $\mathcal{S}_i(l_u,l_w)$ is the value of the match between $l_u$ and $l_w$.

Since the instances are often associated to textual contents (e.g., posts and comments), we consider TF-IDF and cosine similarity measures in order to understand if and how much textual contents are semantically related.

In more detail, let $\{D_1, \ldots D_m\}$ be a set of textual documents and $w_{ij}$ be a word occurring in the document $D_j$ ($j=1, \ldots, m$). The TF-IDF function for $w_{ij}$ is defined as $TF$-$IDF(w_{ij})=TF(w_{ij})*IDF(w_{ij})$, such that: $TF(w_{ij})=\frac{|w_{ij}|}{|D_j|}$ where $|w_{ij}|$ is the length of $w_{ij}$ and $|D_j|$ is the number of words in $D_j$, and: $IDF(w_{ij})=\log \frac{m}{h}$ where $h \leq m$ is the number of documents where $w_{ij}$ occurs.

 Let $V_1$ and $V_2$ be two arrays of $k$ real values. The {\it cosine similarity} between $V_1$ and $V_2$ is defined as: $\mathcal{CS}(V_1,V_2)=\frac{\sum_{i=1}^{k}V_1[i]*V_2[i]}{\sqrt{\sum_{i=1}^k V_1[i]^2} * \sqrt{\sum_{i=1}^k V_2[i]^2}}.$ 
The similarity between two nodes associated to textual contents may be then computed as the cosine similarity between arrays containing the TF-IDF values of the words occurring in those contents.

\subsection{Framework architecture}
\label{sub::architecture}
We describe here the general architecture of a framework implementing our approach, by also discussing different possible choices for its components. In Section \ref{sec::exp} we show how the framework may be specialized for some considered case studies. The framework consists of five steps: data collection, data transformation and integration, network construction, user profiling, profile matching.

\paragraph{Data Collection} 
Data collection is related to the recovery of data generated by the different actors of the system. OSNs such as Facebook or Twitter are among the main sources of data and they can be used to retrieve different types of information related to the users. Other possible sources of information  are web sites, blogs, forums, virtual communities, and review web-sites.
One of the most common technique for data collection from OSN is \textit{web scraping}, which consists in developing a program able to act like a browser by saving the HTML code of the requested web pages \cite{Parvez18}. As will be specified in Section \ref{sec::exp}, we adopt web scraping for this step. However, 
some service providers prevent external users and automatic software from accessing and searching a complete set of information stored in their databases. Therefore, suitable web applications are often applied for data collection by enabling users to specify their virtual profiles and related information.
In addition, Application Programming Interfaces (APIs) exposed by different OSN's services (such as, the Graph API of Facebook, or the Web API of Spotify) can be used in order to retrieve the social data of the users by requesting them directly to the OSN's provider. 

\paragraph{Data Transformation and Integration}
Data collected from different accounts and social media need to be first normalized in order to proceed with the further step of integration. Then, data will be cleaned thus that errors and inconsistencies are removed. Several automatic tools are freely available for data cleaning and integration (e.g., Kettle \cite{kettle_book}). This task provides as result complete and consistent information, that will be used for the creation of network models and profiles in the following two steps.

\paragraph{Network Construction}
Information on both users and their relationships (e.g., friendship on OSNs) collected from social media, transformated and integrated in the previous steps, is then modeled as networks. A network here is represented by an undirect graph where nodes are associated to users, and edges to their relationships. Both nodes and edges are labeled: nodes are associated to identifiers of user profiles, and edge labels indicate the type of relationship and/or its strength. Depending on the applications that will be run on the so built networks, they will be stored in suitable databases (e.g., ArangoDB \cite{arangodb}) and formats (e.g., JSON).

\paragraph{User Profiling}
Suitable user profiles are built which complement network information. That is, each node in the network points to a tree, as explained in Section \ref{sub::basics}, where data associated to a user and retrieved from the considered social media are hierarchically stored in the form of categories, sub-categories and their specific instances. In more details, when users register to a DOSN service, they specify personal information such as the user's full
name, the languages that user knows, age, birthday, gender, email, the relationship status, political view and the religion. Further information comes from the current and home town location, usually including city, country, state, latitude, longitude, street and postcode. Education and work experiences, again with the corresponding locations, are stored as further categories. Moreover, users may  specify information about their general interests, such as music, sport and movies. 
User friendships (referred as Friends), group memberships and properties of relationships (e.g., strength, trust, etc.) are taken into account by the Friends sub-categories in the user's profile. Moreover, another type of information comes from private communications, posts, comments, short text messages that authorized users can write and attach to the content (i.e. posts or images), each organized in suitable sub-trees on the user profile. 

\paragraph{Profile Matching}
Let $P_u$ and $P_w$ be two user profiles and, with references to the definition of match provided in Section \ref{sub::basics}, let $\{\mathcal{S}_1, \ldots, \mathcal{S}_k\}$ and $\{th_1, \ldots, th_k\}$ be the similarity functions and thresholds defined for the $k$ categories in the input dataset. We compute the match between $P_u$ and $P_w$ according to the following procedure. For each category $i$, if the corresponding node is present in both $P_u$ and $P_w$ then proceed with the comparison of the two sub-trees $P_u^i$ and $P_w^i$ rooted by the node associated to $i$. For each sub-category $h$ of $i$, if the corresponding node is present in both the two profiles then compare the two sub-trees rooted by $h$, and continue recursively until the leaves are reached.  The value of the match between $P_u^i$ and $P_w^i$ is given by: $\mu_i(P_u,P_w)=\frac{\sum_{j=1}^{h}\mathcal{S}_{i}^j(l_u,l_w)}{h}$, where $h$ is the number of leaf pairs $l_u^i$ and $l_w^i$ in $P_u^i$ and $P_w^i$, respectively, such that $\mathcal{S}_i(l_u^i,l_w^i) \geq th_i$. The final match between the two profiles is computed as: $\mu(P_u,P_w)=\frac{\sum_{i=1}^{k'} \mu_i(P_u,P_w)}{k'}$, if $k'$ is the number of common categories between $P_u$ and $P_w$. 

It is worth pointing out that, with this definition of profile matching, two different profiles may present a good match value also when entire portions of them do not match. This is due to the fact that only the subtrees corresponding to categories and sub-categories that are present in both profiles are considered. However, the final goal of our approach is to identify the most convenient users for advertisements spreading, given a specific brand and a specific advertisement campaign. Therefore, the fact that entire portions of the user profiles may be discarded is not relevant to this aim.

\section{Experimental Validation}
\label{sec::exp}
\begin{table}[h]
\caption{Facebook Pages}
\vspace{-0.5cm}
\begin{center}
\footnotesize{
\begin{tabular}{|c|c|c|c|}
\hline
\textbf{Category}&\textbf{Name}& \textbf{Topics}& \textbf{\#Follows$^{\mathrm{a}}$} \\
\hline
company&@BarillaIT&food products&2.7M  \\
company& @AudiIT&road vehicles&1.7M\\
company& @AlfaRomeoIT&road vehicles&2.8M\\
company& @Inter&football club&12M\\
company& @Juventusfcita&football club&37M\\
company&@KikoMilanoIT&cosmetic &4.5M\\
company&@Carpisa.it&suitcases, bags&910K\\
community&@BlogGiallozafferano&cooking recipes&2.2M  \\
community& @CucinaFanpage.it&cooking&4.6M\\
community&@ilrompipallone&football&1.2M\\
community&@GliAutogol&football&2M\\
community&@studenteincrisi.official&students&912K\\
community&@cepostaperteufficiale&TV programme&1.3M\\
community&@tomshardware.it&technology&192K\\
public figure&@SudSoundSystemOfficial&band, music&90K\\
public figure&@minamazziniofficial&singer, music&1.1M\\
public figure&@fedezofficial&singer, music&2.2M\\
\hline
\multicolumn{4}{l}{$^{\mathrm{a}}$with reference to the year of the data collection 2018.}
\end{tabular}
}
\label{tab1}
\end{center}
\vspace{-0.5cm}
\end{table}
We have applied our framework to real data taken from Facebook. We have configured the framework in order to collect relevant information on a set of Facebook {\it (i)} users and  {\it (ii)} pages representing communities, public figures, companies, brands. We have simulated some advertising campaigns in order to identify the set of users who could be the best targets, based on the information obtained from their profiles.

\paragraph{Data Collection}
Web scraping has been used in order to collect and extract both the posts published from different Facebook pages, and the profiles information and posts published by a set of Facebook users. Table \ref{tab1} shows the selected Facebook pages and their type, name, topics and total number of follows. We have collected a total of $18$ Facebook pages including companies ($39\%$), communities ($39\%$), and public figures ($22\%$). As shown from the total number of follows, the resulting Facebook pages are very popular and they are related to different areas of interest, such as food products, cooking, vehicles, football, technology, cosmetics, television, school, and music. As for the collected users, the resulting dataset is quite heterogeneous and it consists of $345$ Facebook users: $202$ males and $143$ females, with average age of $25.3$. Only $310$ users specify information about their geographic location, while the average number of posts published by those users is equal to $162$.

\paragraph{Data Transformation and Integration}
We have implemented suitable scripts in order to resolve data inconsistency, posts without text content, and duplicates on the collected data.

\paragraph{Network Construction}
The network associated to the considered Facebook profiles has been built, starting from three different access points.

\begin{figure}[tb]
\centering 
\subfigure[Text cloud Companies\label{fig:textCloudCompany}]{
        \includegraphics[width=0.30\textwidth]{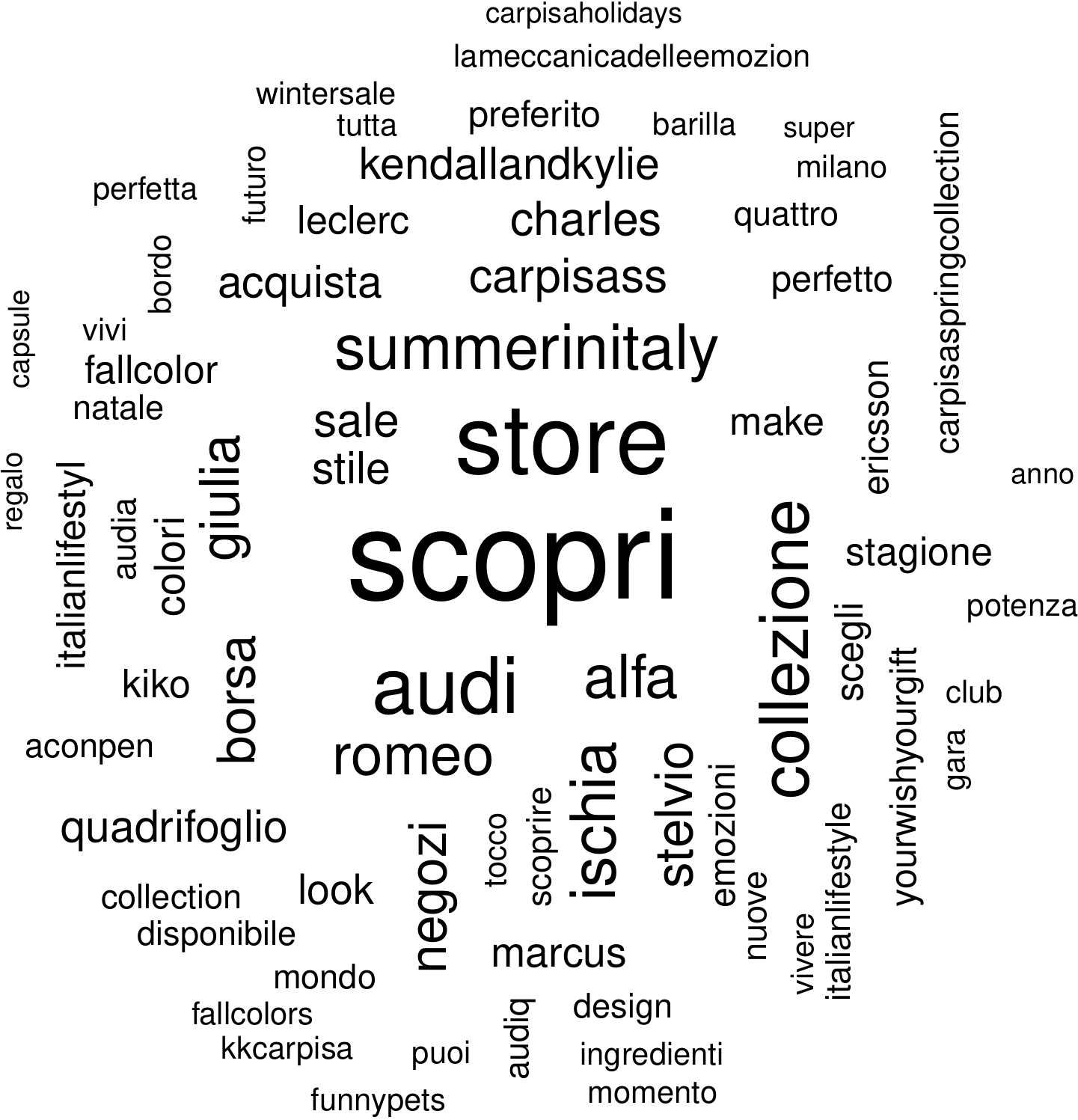}}
\subfigure[Text cloud Communities\label{fig:textCloudCommunity}]{
        \includegraphics[width=0.30\textwidth]{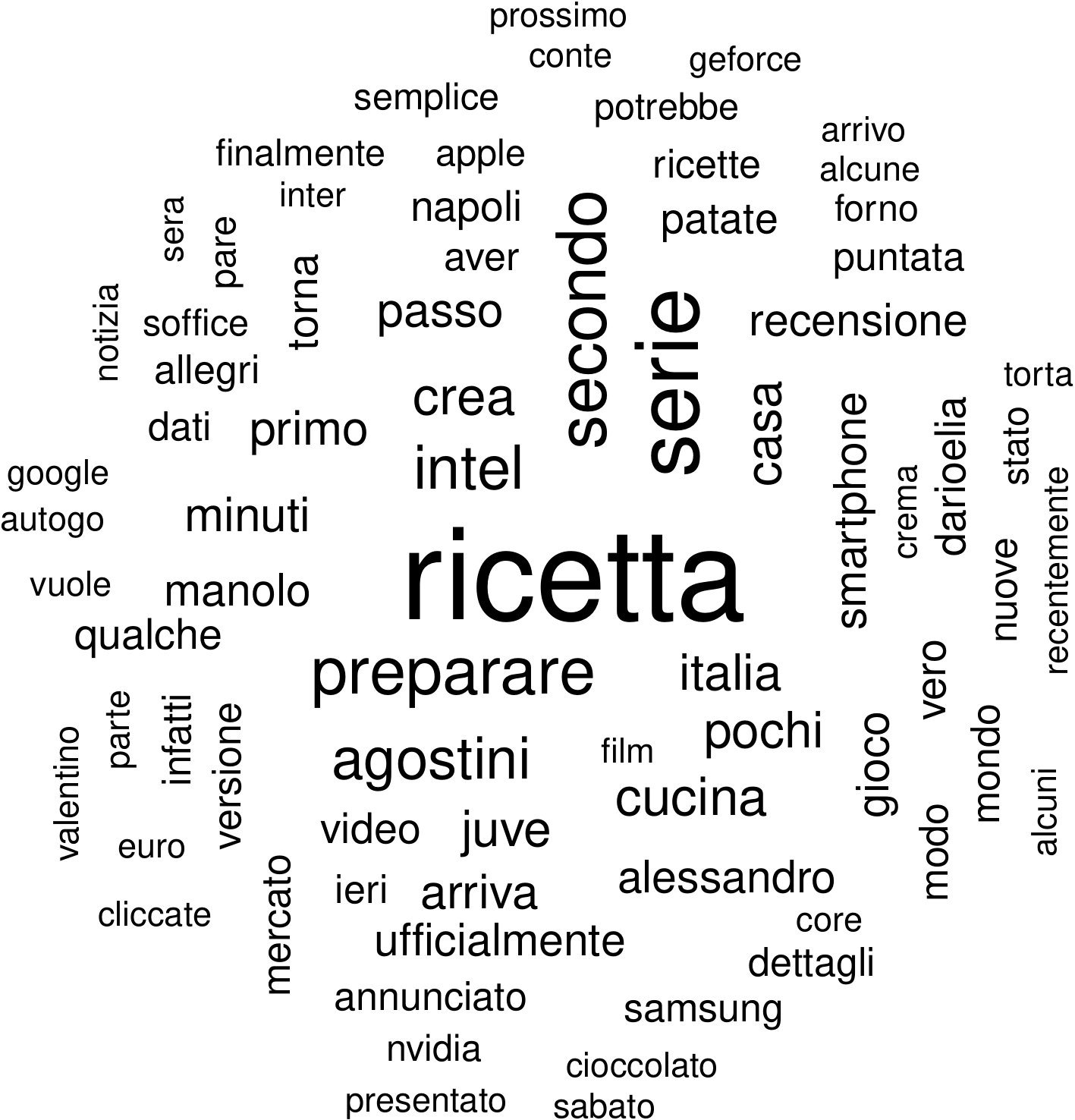}}
\subfigure[Text cloud Public figures\label{fig:textCloudFigure}]{
        \includegraphics[width=0.30\textwidth]{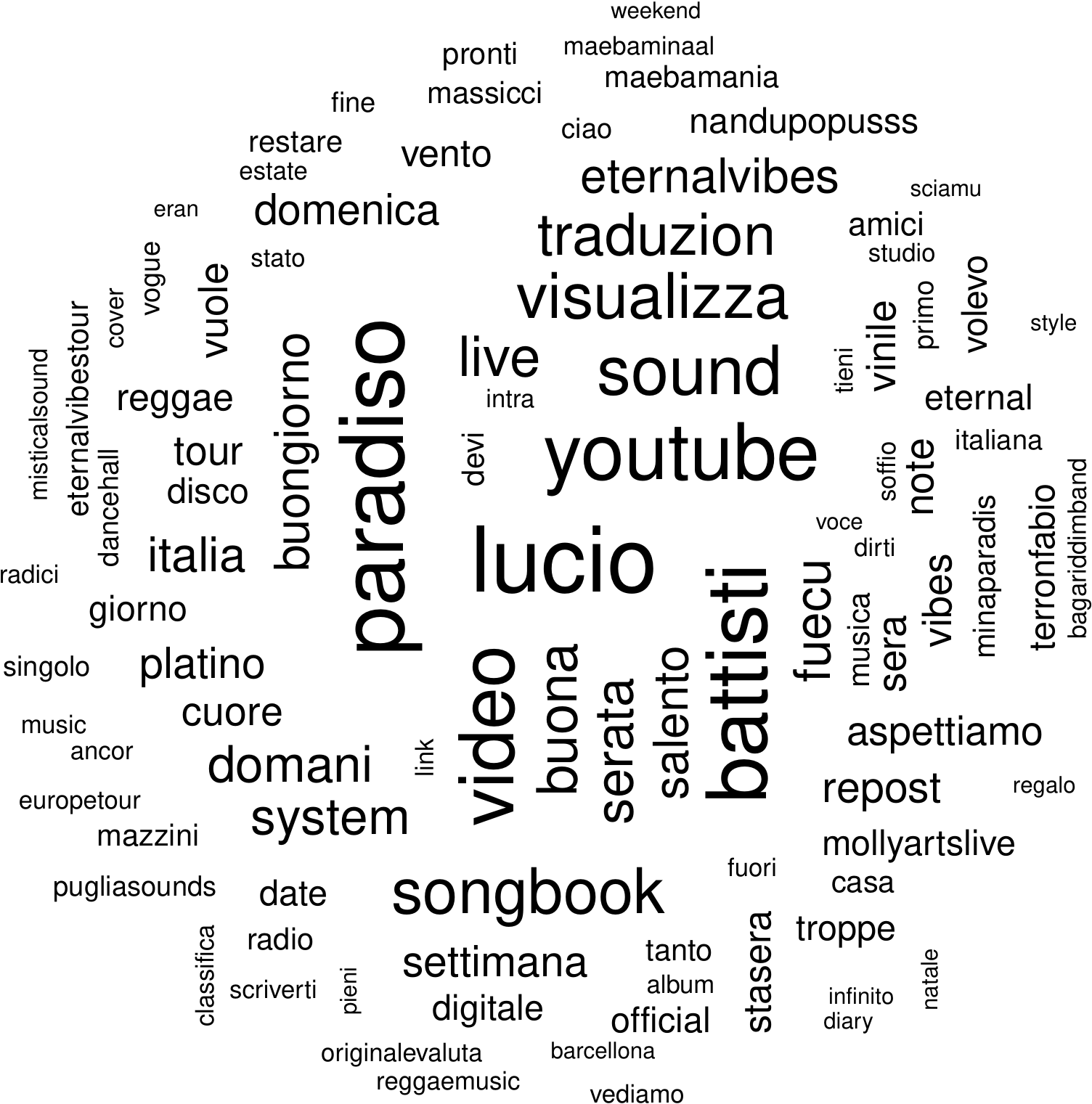}}
\caption{Statistics about the most frequent terms used by Facebook pages.}
\label{fig:Textmining}
\vspace{-0.5cm}
\end{figure}

\paragraph{User Profiling}
For both Facebook users and pages we have created profiles according to the procedure described in Section \ref{sub::architecture}. In particular, in our analysis we focus on gender, age, education, job and posts associated to the profiles. Figures \ref{fig:textCloudCompany}, \ref{fig:textCloudCommunity}, and \ref{fig:textCloudFigure} show the first $100$ prominent terms for pages of the three considered types (the size of each term is proportional to its frequency of occurrence). For example, \textit{store} is one of the most used words in the case of posts published by companies' Facebook pages. Furthermore, there are also a high occurrence of terms which are related either to the brand name (such as \textit{Audi}) or to the best-selling models. Instead, in the case of communities' Facebook pages, the most frequent terms used in the posts mainly depends on the topics discussed by the corresponding community. For example, among the most frequent terms we have \textit{receipts}, \textit{intel}, and \textit{juve} because the majority of the pages discuss food, technology, and sport topics.
Finally, the posts published by the pages of public figures deal with music interests since they are associated with known singers and bands. 

\paragraph{Profile Matching}
\vspace{-0.3cm}
\begin{figure}[h]
\centering 
        \includegraphics[width=0.90\textwidth]{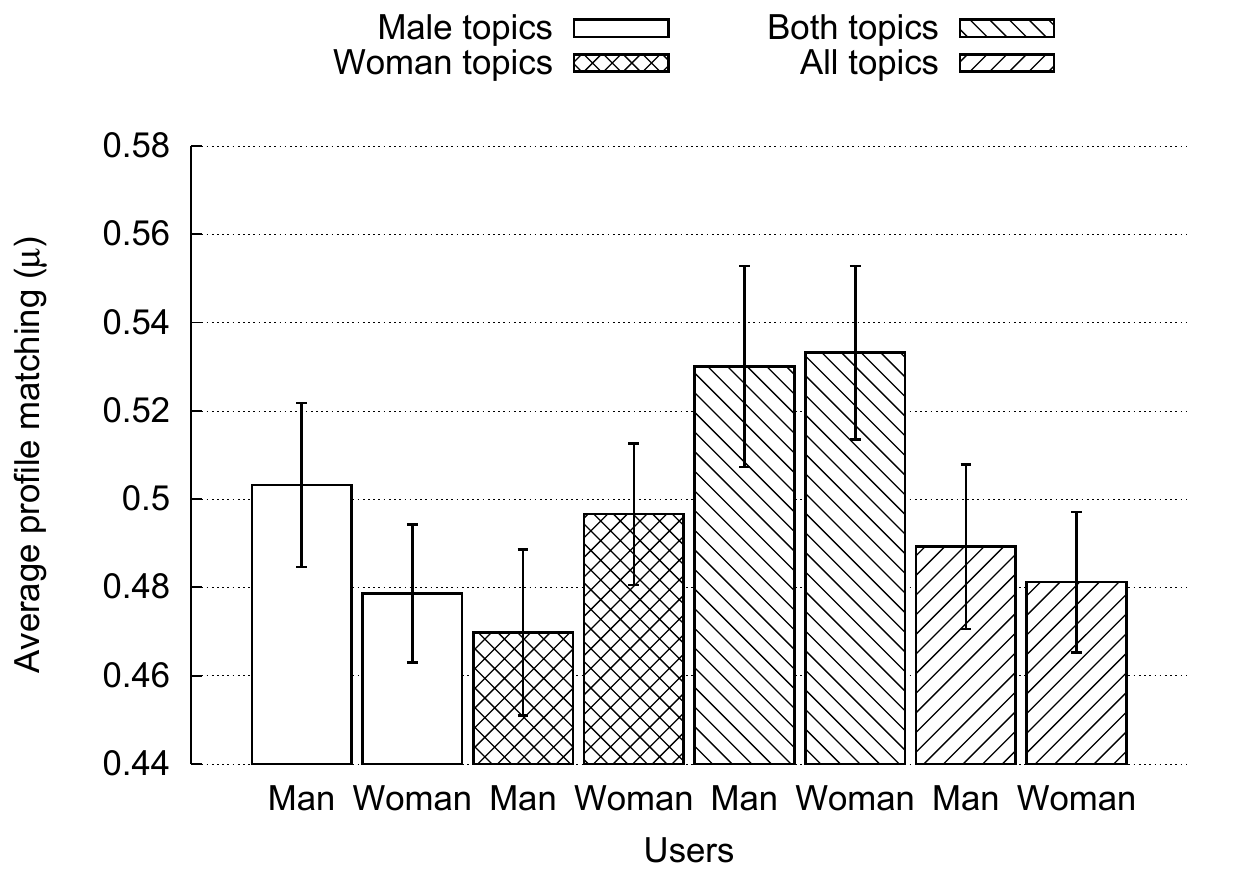}
\caption{Average match between the profiles of Facebook pages and users.}

\label{fig:simmTop}
\end{figure}
%

We have first applied the proposed framework on a very simplified case, where the only considered category for profile matching is posts. The cosine similarity recalled in Section \ref{sub::basics} has been applied for the match computation, after that all posts of a specific user (or page) have been collected inside a document and the bags of words based on TF-IDF are considered. We have classified the profiles corresponding to Facebook pages according to the topics of interest, by distinguishing those topics that could be of great interest for women, men, and both genders. As an example, female users may be more interested to Facebook pages that discuss topics related to cosmetics and handbags, whereas male users to road vehicles and football. No gender is preferential for those pages related to students, music, and singers.
Figure \ref{fig:simmTop} shows the average match between the profiles of Facebook men/women users and the profiles of Facebook pages, with respect to the topics discussed in the different pages, suitably classified in \textit{Man topics}, \textit{Woman topics}, \textit{Both topics}. The average match w.r.t. all Facebook pages in our dataset is included as well (\textit{All topics}). Results show that the framework correctly classifies users w.r.t. the match between their interests and the contents in the (brand) pages.

We have then simulated three different simple advertising campaigns, and searched for the most suitable users to be used as target for each campaign. We have considered gender, age, education, job and posts associated to the profiles. In particular, for each campaign we have computed the match between the considered brand profile and all users, then sorted users according to the resulting match value and selected the top $3\%$ users. The first example refers to the advertisement of ``pasta'', from {\it BarillaIT}. As expected, the resulting target users are various with respect to all considered categories. The second experiment refers to female handbags produced by {\it Carpisa}, and the recommended target users include only women. Finally, two different singers have been considered, the first one popular among teen-agers and the second one among thirty-five. Again, the suggested target users correspond exactly to the right category, that is age group in this case.

\section{Concluding Remarks}
Experimental evaluations show that our framework is able to correctly identify a set of suitable users to be proposed as possible consumers in specific advertisement campaigns. The framework can be extended according to several directions. Notably among them, the integration of data coming from different OSN (e.g., Facebook, Twitter, Instagram), the inclusion of a larger number of categories in the profiling and profile matching steps, the adoption of big-data technologies for the storage and management of larger input datasets.

\section*{Acknowledgments}
Partially supported by the Italian Ministery of University and Research under the grant PRIN 2017WR7SHH ``Multicriteria Data Structures and Algorithms: from compressed to learned indexes, and beyond''.

\bibliographystyle{plain}
\bibliography{biblio}

\end{document}